\documentclass[conference]{IEEEtran}
\usepackage{cite}
\usepackage{amsmath,amssymb,amsfonts}

\usepackage{algorithmic}
\usepackage{subfigure}

\usepackage{graphicx}
\graphicspath{{./figures/}}

\newcommand{\be}{\begin{equation}}
\newcommand{\ee}{\end{equation}}

\usepackage[bookmarks=true]{hyperref}
\hypersetup
{
    colorlinks=true, 
    linktoc=all,     
    linkcolor=black, 
    allcolors=black, 
    pdfstartview=, 
    pdfremotestartview=, 
}
\usepackage[all]{hypcap}
\usepackage{bookmark}
\usepackage{cleveref}

\def\BibTeX{{\rm B\kern-.05em{\sc i\kern-.025em b}\kern-.08em
    T\kern-.1667em\lower.7ex\hbox{E}\kern-.125emX}}

\begin{document}

\title{On Deep Learning Classification of Digitally Modulated Signals Using Raw I/Q Data}

\author{\IEEEauthorblockN{John A. Snoap and Dimitrie C. Popescu}
\IEEEauthorblockA{ECE Department, Old Dominion University \\
Norfolk, VA 23529, USA \\
\{jsnoa001, dpopescu\}@odu.edu \vspace{-0.5cm} }
\and
\IEEEauthorblockN{Chad M. Spooner}
\IEEEauthorblockA{NorthWest Research Associates \\
Monterey, CA 93940, USA \\
cmspooner@nwra.com \vspace{-0.5cm} }
}

\maketitle

\begin{abstract}
The paper considers the problem of deep-learning-based classification of digitally modulated signals using I/Q data and studies the
generalization ability of a trained neural network (NN) to correctly classify digitally modulated signals it has been trained to recognize
when the training and testing datasets are distinct. Specifically, we consider both a residual network (RN) and a convolutional neural
network (CNN) and use them in conjunction with two different datasets that contain similar classes of digitally modulated signals but
that have been generated independently using different means, with one dataset used for training and the other one for testing.
\end{abstract}


\vspace{-0.1cm}
\section{Introduction}\label{sec:Intro}
Identifying a specific digital modulation scheme without specific knowledge about its parameters and/or without prior information
on the transmitted data can be accomplished using likelihood-based approaches \cite{Xu_etal_TSMC2011, Hameed_etal_TW2009}
or cyclostationary signal processing (CSP) \cite{Spooner_etal_Asilomar2000}. Recently, deep learning (DL) approaches to classifying
digitally modulated signals have also been proposed, such as those in \cite{Tim2018, Tim2017}, which use the raw I/Q signal components
for training and signal recognition/classification, or the alternative approaches in \cite{Sun2018, Zhang2020, Kulin2018, Rajendran2018, Bu2020},
in which the amplitude/phase or frequency domain representations are used. 

DL-based approaches implement neural networks (NNs) and rely on their extensive training to make the distinction among different
classes of digitally modulated signals. Usually, the available data set is partitioned into subsets that are used for training/validation
and testing, respectively, and it is not known in general how well NNs trained on a specific dataset respond to similar signals that are
generated differently than those in the dataset used for training and validation. However, the problem of the dataset shift, also referred
to as out-of-distribution generalization \cite{Djolonga_etal_arXiv2020}, is an important problem in machine-learning-based approaches,
which implies that the training and testing data sets are distinct, and that data from the testing environment is not used for training the
classifier. This problem motivates the work presented in this paper, which studies the performance of DL-based classification of digitally
modulated signals when the training and testing data are taken from distinct data sets.

Specifically, the paper considers the DL-based approach to modulation classification discussed in \cite{Tim2017, Tim2018}, which uses
raw I/Q signal data to train NNs for use in classifying digitally modulated signals. To assess the generalizability of I/Q-trained NNs and
evaluate how well they perform on classifying digitally modulated signals from datasets they were not trained on, we implement NNs
with architectures similar to those in \cite{Tim2018} training them with signals from one data set and evaluating their classification
performance using signals in the other data set. The two data sets used are publicly available:
\begin{itemize}
\item \texttt{DataSet1} was generated for use in \cite{Tim2018} and can be downloaded from DeepSig Inc.'s website \cite{DeepSigDataSet}.
\item \texttt{DataSet2} was independently generated and is available on the CSP Blog \cite{MLChallenge}.
\end{itemize}
\vspace{-0.05cm}

The paper is organized as follows: Section~\ref{sec:NNModels} describes the NNs implemented for classification of digitally modulated
signals, followed by presentation of the two data sets used for training and testing in Section~\ref{sec:Datasets}. Section~\ref{sec:NumericalResults}
discusses NN training details and presents numerical results showing the classification performance of the two NNs, with a focus on
generalization, when training and testing are performed using signals from different data sets. The paper is concluded with final remarks
in Section~\ref{sec:Conclusion}.

\vspace{-0.2cm}
\section{Neural Network Models for Digital Modulation Classification}\label{sec:NNModels}
\vspace{-0.1cm}
DL NNs consist of multiple interconnected layers of neuron units and include an input layer, which takes the available data for processing,
several hidden layers that provide various levels of abstractions of the input data, and an output layer, which determines the final classification
of the input data \cite{Goodfellow-et-al-2016}. The hidden layers of a NN include:
\begin{itemize}
\item Convolutional layers, may be followed by batch normalization to increase regularization and avoid overfitting.
\item Fully connected layers that may be preceded by dropout layers.
\item Nonlinear layers, with the two common types of nonlinearities employed being the rectified linear unit (ReLU) and the scaled exponential
linear unit (SELU).
\item Pooling layers that use average or maximum pooling to provide invariance to local translation.
\item A final softmax layer that establishes the conditional probabilities for input data classification by determining the output unit activation function
for multi-class classification problems.
\end{itemize}

\begin{table}
\centering
\caption{Residual Neural Network Layout}
{
\vspace{-0.25cm}
\begin{tabular}{ c c }
\hline\hline
Layer & Output Dimensions \\
\hline
Input & $2 \times 1024$ \\
Residual Stack & $32 \times 512$ \\
Residual Stack & $32 \times 256$ \\
Residual Stack & $32 \times 128$ \\
Residual Stack & $32 \times 64$ \\
Residual Stack & $32 \times 32$ \\
Residual Stack & $32 \times 16$ \\
Drop($50\%$)/FC/SELU & $128$ \\
Drop($50\%$)/FC/SELU & $128$ \\
Drop($50\%$)/FC/SoftMax & \# Classes \\
\hline\hline
\end{tabular}
}\label{table:RNnetwork}
\end{table}

\begin{table}
\vspace{0.1cm}
\centering
\caption{Residual Stack Layout}
{
\vspace{-0.25cm}
\begin{tabular}{ c c }
\hline\hline
Layer & Output Dimensions \\
\hline
Input & $X \times Y$ \\
$1 \times 1$ Conv & $32 \times Y$ \\
Batch Normalization & $32 \times Y$ \\
ReLU & $32 \times Y$ \\
Residual Unit & $32 \times Y$ \\
Residual Unit & $32 \times Y$ \\
Maximum Pooling & $32 \times Y/2$ \\
\hline\hline
\end{tabular}
}\label{table:ResStack}
\end{table}
\begin{table}[h!]
\centering
\caption{Residual Unit Layout}
{
\vspace{-0.25cm}
\begin{tabular}{ c c }
\hline\hline
Layer & Output Dimensions \\
\hline
Input & $X \times Y$ \\
Conv & $32 \times Y$ \\
Batch Normalization & $32 \times Y$ \\
ReLU\_1 & $32 \times Y$ \\
Conv & $32 \times Y$ \\
Batch Normalization & $32 \times Y$ \\
ReLU\_2 & $32 \times Y$ \\
Addition(Input, ReLU\_2) & $32 \times Y$ \\
\hline\hline
\end{tabular}
}\label{table:ResUnit}
\end{table}
\begin{table}[h!]
\centering
\caption{Convolutional Neural Network Layout}
{
\vspace{-0.25cm}
\begin{tabular}{ c c }
\hline\hline
Layer & Output Dimensions \\
\hline
Input					& $2 \times 1024$ \\
Conv					& $16 \times 1024$ \\
Batch Normalization		& $16 \times 1024$ \\
ReLU					& $16 \times 1024$ \\
Maximum Pooling			& $16 \times 512$ \\
Conv					& $24 \times 512$ \\
Batch Normalization		& $24 \times 512$ \\
ReLU					& $24 \times 512$ \\
Maximum Pooling			& $24 \times 256$ \\
Conv					& $32 \times 256$ \\
Batch Normalization		& $32 \times 256$ \\
ReLU					& $32 \times 256$ \\
Maximum Pooling			& $32 \times 128$ \\
Conv					& $48 \times 128$ \\
Batch Normalization		& $48 \times 128$ \\
ReLU					& $48 \times 128$ \\
Maximum Pooling			& $48 \times 64$ \\
Conv					& $64 \times 64$ \\
Batch Normalization		& $64 \times 64$ \\
ReLU					& $64 \times 64$ \\
Maximum Pooling			& $64 \times 32$ \\
Conv					& $96 \times 32$ \\
Batch Normalization		& $96 \times 32$ \\
ReLU					& $96 \times 32$ \\
Average Pooling			& $96$ \\
Drop($0\%$)/FC/SoftMax	& \# Classes \\
\hline\hline
\end{tabular}
\vspace{-0.25cm}
}\label{table:CNNnetwork}
\end{table}

We note that commonly used types of NNs in DL include CNNs and RNs, with the latter including bypass connections between layers
that enable features to operate at multiple scales and depths throughout the NN \cite{Goodfellow-et-al-2016}.

The study on DL-based classification of digitally modulated signals presented in this paper aims at assessing the generalization ability of
trained NNs and demonstrating the need for more robust testing of NNs trained to recognize digital modulation schemes. In this direction,
we consider both an RN with the layout specified in Tables~\ref{table:RNnetwork}--\ref{table:ResUnit}, and a CNN with the layout specified
in Table~\ref{table:CNNnetwork}. For both types of NNs (RN and CNN) the convolutional layers use a filter size of $23$ as increasing the
filter size increases the computational cost with no performance gain, while decreasing it lowered performance. 

The NN structures outlined in Tables I -- IV are similar to those in~\cite{Tim2018}, and that the novelty of this work consists in the use
of two distinct datasets for training and testing, respectively, which will allow us to evaluate if the NNs are able to generalize and continue
to distinguish the classes of digitally modulated signals for which they have been trained when tested on signals that are generated
differently than the signals in the training dataset.

\section{Datasets for NN Training and Testing the Classification of Digitally Modulated Signals}\label{sec:Datasets}
The DL NNs used for classification of digitally modulated signals with the structure outlined in Section~\ref{sec:NNModels}
are trained and tested using digitally modulated signals in two distinct datasets that are publicly available for general use, which are
referred to as \texttt{DataSet1}~\cite{DeepSigDataSet} and \texttt{DataSet2}~\cite{MLChallenge}, respectively. These datasets contain
collections of the I/Q data corresponding to signals generated using common digital modulation schemes that include BPSK, QPSK,
8-PSK, 16-QAM, 64-QAM, and 256-QAM, with different signal-to-noise ratios (SNRs). Brief details on the signal generation methods
are also provided in the descriptions of the two datasets, and we note that the signals include overlapping excess bandwidths and the
use of square-root raised-cosine (SRRC) pulse shaping with similar roll-off parameters.

According to  \cite{Tim2018}, \texttt{DataSet1} includes both simulated signals and signals captured over-the-air, with $24$ different
modulation types and $26$ distinct SNR values for each modulation type, ranging from $-20$~dB to $30$~dB in $2$~dB increments.
For each modulation type and SNR value combination, there are $4,096$ signals included in the dataset, with $1,024$ I/Q samples
for each signal, and the total number of signals in \texttt{DataSet1} is $2,555,904$.

\texttt{DataSet2} \cite{MLChallenge} contains only simulated signals corresponding to $8$~different digital modulation types with SNRs
varying between $0$~dB and $13$~dB. The dataset includes a total of $112,000$ signals, with $32,768$ I/Q samples for each signal.

One key difference between \texttt{DataSet1} and \texttt{DataSet2} is that in the former the total SNR (i.e., the ratio of the signal power
to the total noise power within the sampling bandwidth) is used, while in the latter the in-band SNR (i.e., the ratio of the signal power to
the power of the noise falling within the signal's actual bandwidth) is used.
Because the listed SNRs for the signals in \texttt{DataSet2} correspond to in-band SNR values, we have also estimated the total SNR
corresponding to the sampling bandwidth for signals in \texttt{DataSet2}. This turned out to be about $8$~dB below the in-band SNRs
and is useful for a side-by-side comparison with the signals in \texttt{DataSet1}.

We note that, although the two datasets considered contain common modulation types, the signals in \texttt{DataSet1} have been generated
independently and by different means than the signals in \texttt{DataSet2}, which implies that they are well suited for our study to assess the
out-of-distribution generalization ability of a trained NN to classify digitally modulated signals.

\section{NN Training and Numerical Results}\label{sec:NumericalResults}
Numerical results have been obtained by training both the RN and CNN twice, first using \texttt{DataSet1} then again separately
using \texttt{DataSet2}. We note that, to make the signals in \texttt{DataSet2} compatible with a DL NN trained
on \texttt{DataSet1}, we split each signal in \texttt{DataSet2} into $32$ separate signals with the same class labels to both increase
the total number of signals and reduce the samples per signal to $1,024$, which yielded a total of $3,584,000$ digitally modulated
signals, each with $1,024$ samples, similar to the ones in \texttt{DataSet1}.

For each training, the datasets have been divided into subsets that have been used as follows:  $75\%$ of the signals in a given
dataset have been used for training, $12.5\%$ for validation during training, and the remaining $12.5\%$ for testing after training
has completed.
Prior to training, the signals have been normalized to unit power, and the SGDM algorithm has been implemented for training with a
mini-batch size of $256$ training signals.  Twelve training epochs were used, as further training beyond 12~epochs did not appear to
result in significant performance improvement.

The NNs have been implemented in MATLAB and trained on a high-performance computing cluster with 21 graphical processing unit
(GPU) nodes available, consisting of Nvidia K40, K80, P100, and V100 GPU(s) with 128 GB of memory per node.  We note that
NN training is computationally intensive and takes about 34 hours to complete training for one NN on a single Nvidia K80 GPU.

\begin{figure}
\begin{center}
\subfigure[Training and testing using \texttt{DataSet1}]
{\includegraphics[scale=0.35]{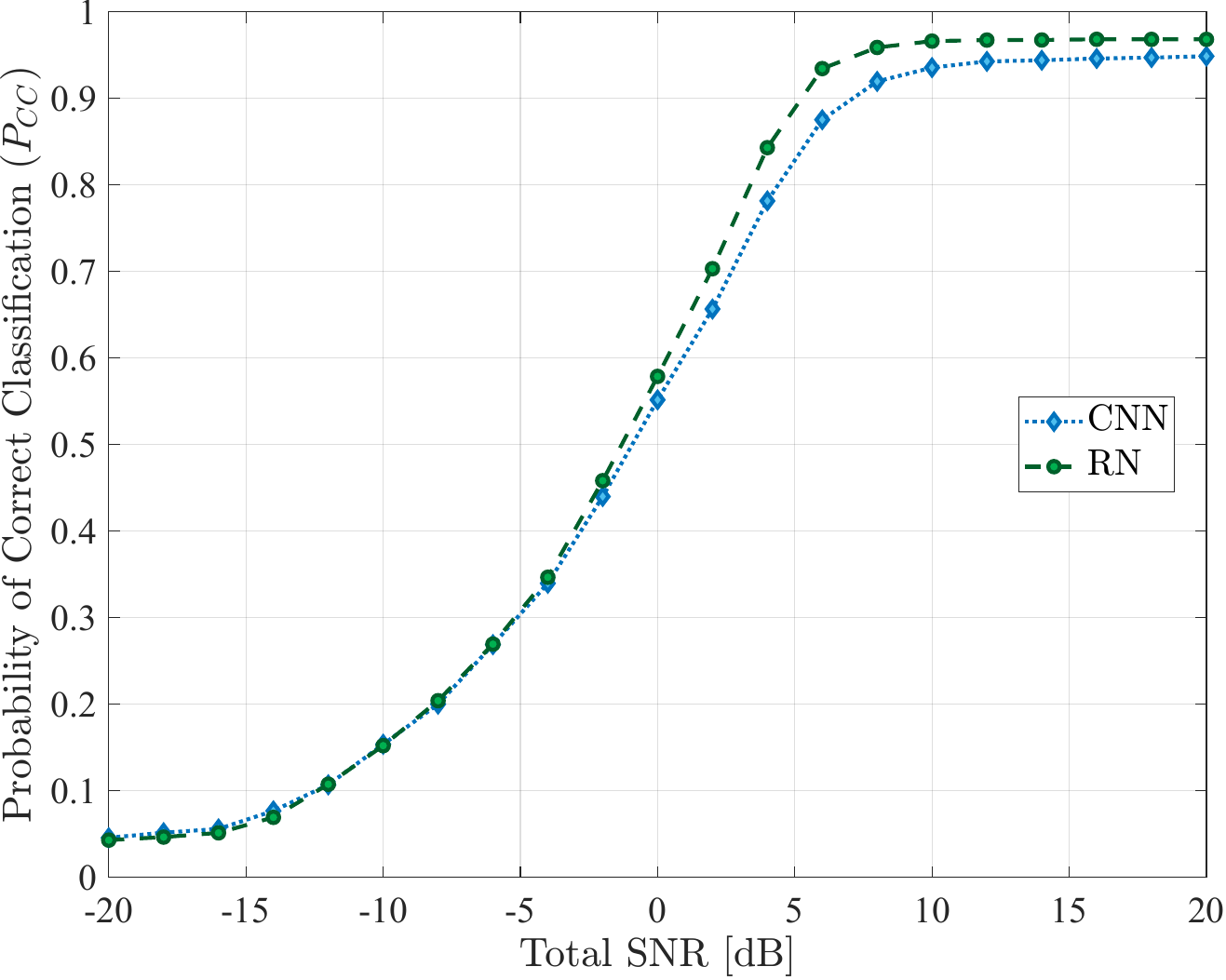}}
\subfigure[Training and testing using \texttt{DataSet2}.]
{\includegraphics[scale=0.35]{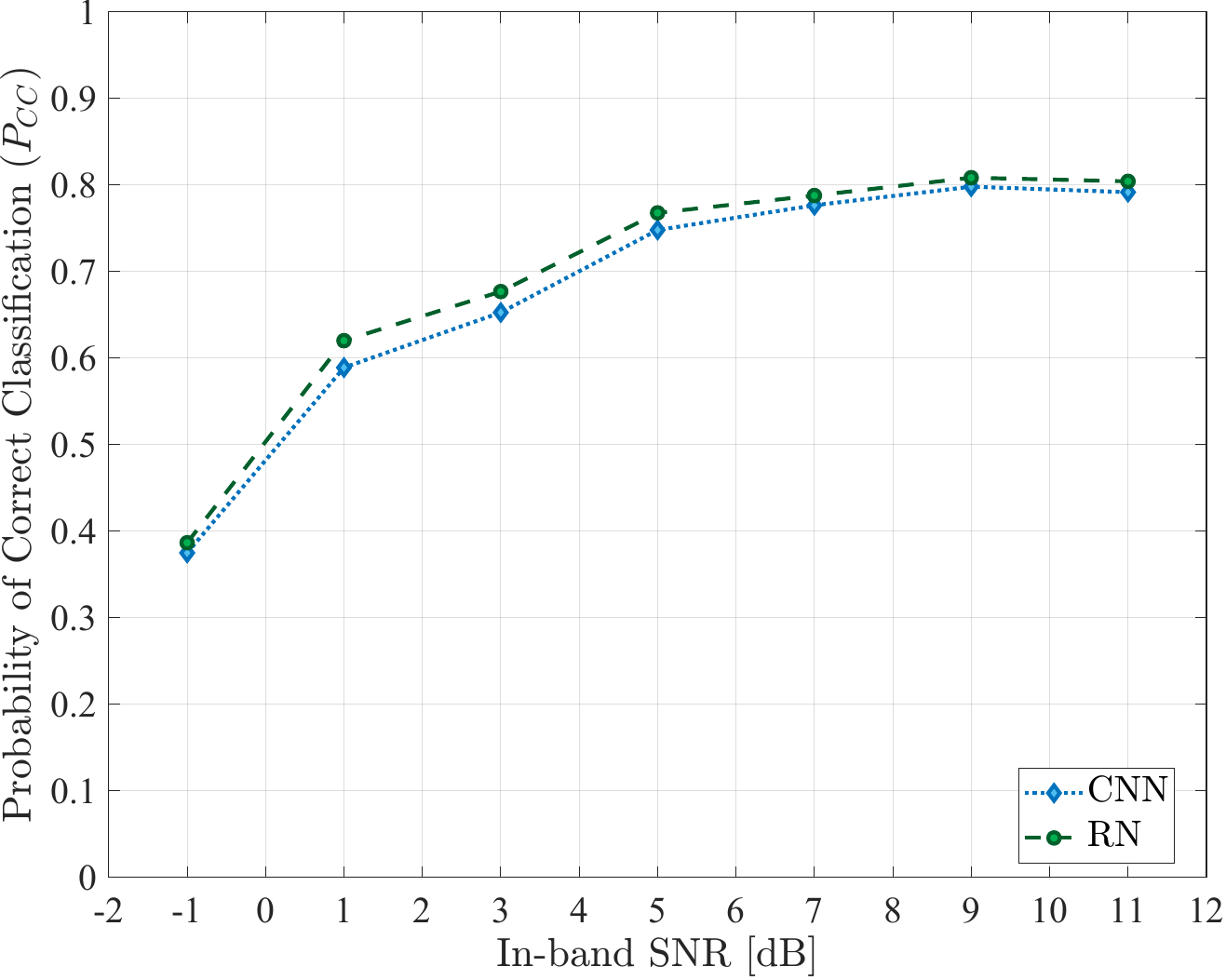}}
\vspace{-0.25cm}
\caption{Classification performance of trained NNs when training and testing signals come from the same dataset.}\label{fig:Train=Test}
\vspace{-0.75cm}
\end{center}
\end{figure}

The first experiment performed involves evaluating the performance of the trained RN and CNN using test signals coming from the
same dataset as the signals used for training. Results from this experiment are shown in Fig.~\ref{fig:Train=Test}, from which we
note that, as expected for a signal processing algorithm, the classification performance improves with increasing SNR. We also note that the use of
\texttt{DataSet1} for NN training and testing corresponds to the same scenario considered in \cite{Tim2018}, and the corresponding
plot shown in Fig.~\ref{fig:Train=Test}(a) is similar to the probability of classification plots presented in \cite{Tim2018}. This is
further corroborated by confusion matrix results, which are similar to those in \cite{Tim2018}, but are omitted from the presentation
due to space constraints.

When \texttt{DataSet2} is used for NN training and testing, the corresponding probability of classification plot shown in Fig.~\ref{fig:Train=Test}(b)
levels at a slightly lower value than when \texttt{DataSet1} is used. This is because the two datasets are different, with more random variables in
\texttt{DataSet2} including a larger range of SRRC roll-off values, CFOs, randomized symbol rates, and signal power levels.  Also, the architectures
of the RN and CNN employed were very similar to those considered in \cite{Tim2018} and have been optimized for \texttt{DataSet1}; further
improvement on \texttt{DataSet2} may be possible by changing the NN architecture.

\begin{figure}
\begin{center}
\subfigure[Training using \texttt{DataSet1}, testing using \texttt{DataSet2}.]
{\includegraphics[scale=0.35]{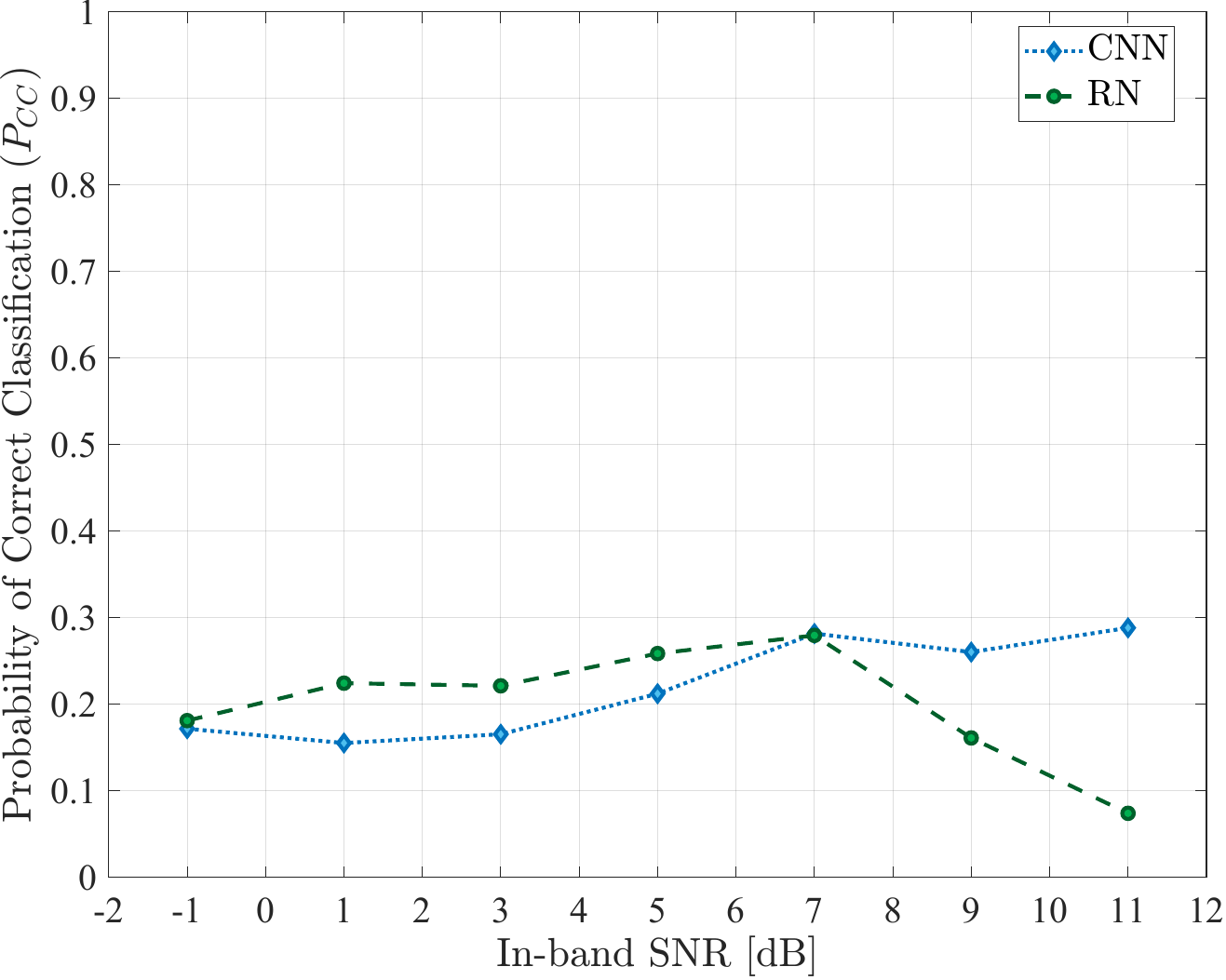}}
\subfigure[Training using \texttt{DataSet2}, testing using \texttt{DataSet1}.]
{\includegraphics[scale=0.35]{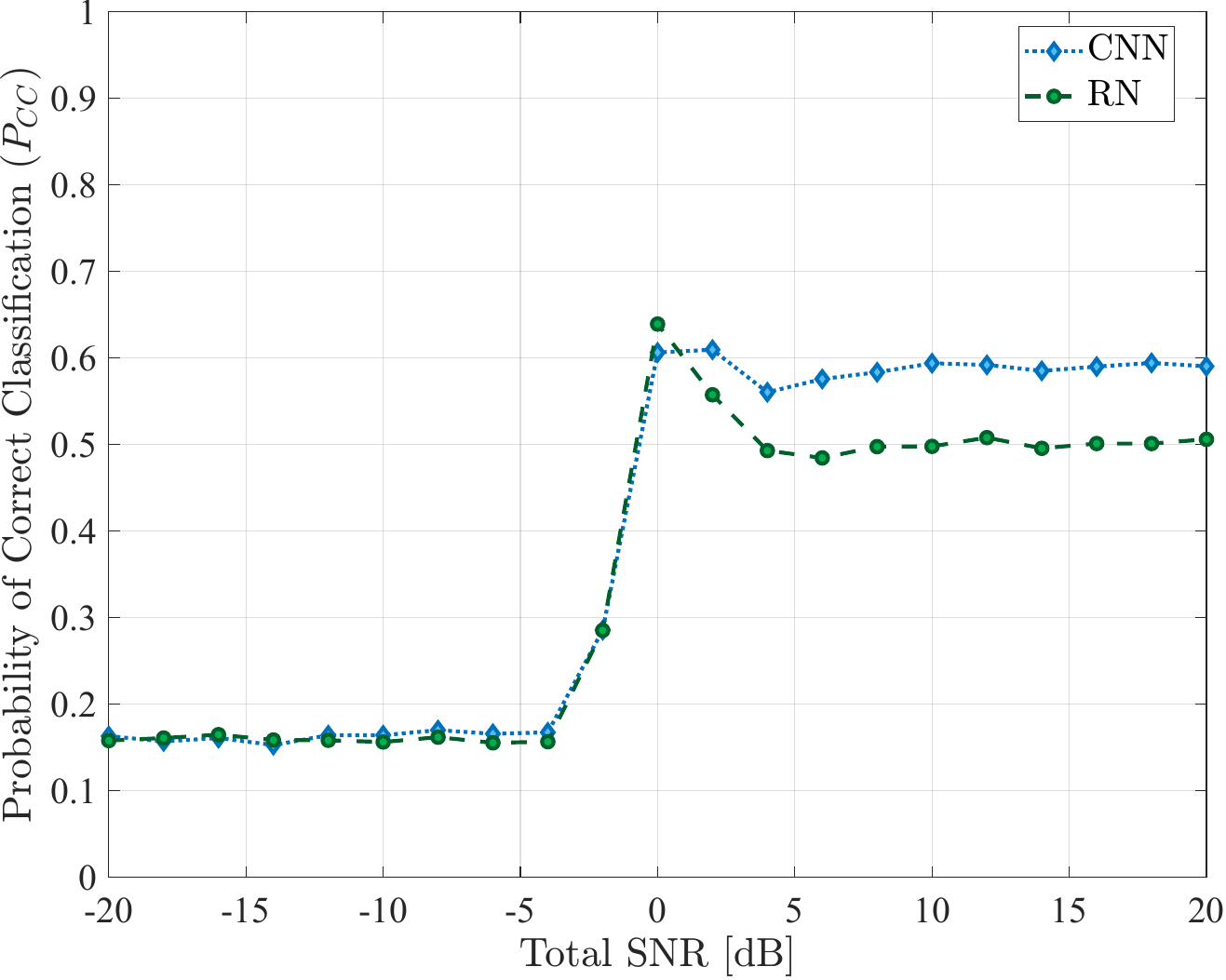}}
\vspace{-0.25cm}
\caption{Classification performance of trained NNs when training and testing signals come from different datasets.}\label{fig:Train!=Test}
\vspace{-0.75cm}
\end{center}
\end{figure}

In the second experiment, we evaluated the performance of trained RN and CNN using test signals coming from a different
dataset than the one used for training. Results from this experiment are shown in Fig.~\ref{fig:Train!=Test}, from which it
is apparent that the generalization performance of the trained NNs is poor, and that they fail to recognize digitally modulated
signals belonging to classes they have been trained on when these signals come from a dataset that is different than the one the
NNs have been trained on. We note that the use of \texttt{DataSet1} for training the NNs does not seem to imply any
generalization abilities as the corresponding probability of correct classification shown in Fig.~\ref{fig:Train!=Test}(a)
is essentially flat around $0.2$ for all SNR values.
While we omit them from the presentation due to space constraints, confusion matrices for this case confirm that the NNs are
not able to generalize and  classify any of the modulation schemes for which they have been trained, when tested with signals
that are not part of the training \texttt{DataSet1}.

When using \texttt{DataSet2} for training the NNs, the corresponding probability of correct classification shown in
Fig.~\ref{fig:Train!=Test}(b) displays a jump around $0$~dB SNR from a low value of $0.2$ leveling at values
around $0.5$ and $0.6$ for the RN and CNN, respectively. Thus, in this case, while the overall classification performance
is below that obtained in the first experiment or reported in \cite{Tim2018}, the trained NNs seem to have some
limited generalization ability, which may also be observed by looking at the corresponding confusion matrices.
While these are omitted due to space constraints, we note that, out of the six modulation types which they have
been trained to recognize, the RN continues to successfully classify $3$ (BPSK, QPSK, and 256-QAM), while
the CNN recognizes $4$ (BPSK, QPSK, 8-PSK, and 256-QAM).

\section{Conclusions}\label{sec:Conclusion}
\vspace{-0.1cm}
The paper considers DL classification of digitally modulated signals using raw I/Q data and studies the generalization
ability of NNs by evaluating their classification performance using signals that come from datasets they were not trained on. 
Results indicate that training a NN to perform modulation classification based on the raw I/Q data causes the NN to learn
peculiarities of a particular training dataset, rather than conveying salient signal aspects that are determined by the underlying
digital modulation scheme.  Thus, more robust NN validation approaches, such as the one presented in this paper, 
should be used to ensure good classification performance.

To improve the generalization ability of NNs in classifying digitally modulated signals, future work will consider training them using
specific signal features that can be extracted from the raw I/Q signal data, such as those based on cyclostationary signal processing.

\vspace{-0.1cm}
\section*{Acknowledgment}
\vspace{-0.1cm}
The authors would like to acknowledge the use of Old Dominion University High-Performance Computing facilities
for obtaining numerical results presented in this work.

\vspace{-0.1cm}



 \bibliographystyle{IEEEtran}
 \bibliography{./bib/refs}

\begin{thebibliography}{10}
\providecommand{\url}[1]{#1}
\csname url@samestyle\endcsname
\providecommand{\newblock}{\relax}
\providecommand{\bibinfo}[2]{#2}
\providecommand{\BIBentrySTDinterwordspacing}{\spaceskip=0pt\relax}
\providecommand{\BIBentryALTinterwordstretchfactor}{4}
\providecommand{\BIBentryALTinterwordspacing}{\spaceskip=\fontdimen2\font plus
\BIBentryALTinterwordstretchfactor\fontdimen3\font minus
  \fontdimen4\font\relax}
\providecommand{\BIBforeignlanguage}[2]{{%
\expandafter\ifx\csname l@#1\endcsname\relax
\typeout{** WARNING: IEEEtran.bst: No hyphenation pattern has been}%
\typeout{** loaded for the language `#1'. Using the pattern for}%
\typeout{** the default language instead.}%
\else
\language=\csname l@#1\endcsname
\fi
#2}}
\providecommand{\BIBdecl}{\relax}
\BIBdecl

\bibitem{Xu_etal_TSMC2011}
J.~L. Xu, W.~Su, and M.~Zhou, ``{Likelihood-Ratio Approaches to Automatic
  Modulation Classification},'' \emph{IEEE Transactions on Systems, Man, and
  Cybernetics -- Part C: Applications and Reviews}, vol.~41, no.~4, pp.
  3072--3108, July 2011.

\bibitem{Hameed_etal_TW2009}
F.~Hameed, O.~A. Dobre, and D.~C. Popescu, ``{On the Likelihood-Based Approach
  to Modulation Classification},'' \emph{IEEE Transactions on Wireless
  Communications}, vol.~8, no.~12, pp. 5884--5892, December 2009.

\bibitem{Spooner_etal_Asilomar2000}
C.~M. Spooner, W.~A. Brown, and G.~K. Yeung, ``{Automatic Radio-Frequency
  Environment Analysis},'' in \emph{Proceedings of the Thirty-Fourth Annual
  Asilomar Conference on Signals, Systems, and Computers}, vol.~2, Monterey,
  CA, October 2000, pp. 1181--1186.

\bibitem{Tim2018}
T.~J. {O’Shea}, T.~{Roy}, and T.~C. {Clancy}, ``{Over-the-Air Deep Learning
  Based Radio Signal Classification},'' \emph{IEEE Journal of Selected Topics
  in Signal Processing}, vol.~12, no.~1, pp. 168--179, 2018.

\bibitem{Tim2017}
T.~{O’Shea} and J.~{Hoydis}, ``{An Introduction to Deep Learning for the
  Physical Layer},'' \emph{IEEE Transactions on Cognitive Communications and
  Networking}, vol.~3, no.~4, pp. 563--575, 2017.

\bibitem{Sun2018}
J.~{Sun}, G.~{Wang}, Z.~{Lin}, S.~G. {Razul}, and X.~{Lai}, ``{Automatic
  Modulation Classification of Cochannel Signals using Deep Learning},'' in
  \emph{Proceedings 23rd IEEE International Conference on Digital Signal
  Processing (DSP)}, 2018, pp. 1--5.

\bibitem{Zhang2020}
D.~{Zhang}, W.~{Ding}, C.~{Liu}, H.~{Wang}, and B.~{Zhang}, ``{Modulated
  Autocorrelation Convolution Networks for Automatic Modulation Classification
  Based on Small Sample Set},'' \emph{IEEE Access}, vol.~8, pp.
  27\,097--27\,105, 2020.

\bibitem{Kulin2018}
M.~{Kulin}, T.~{Kazaz}, I.~{Moerman}, and E.~{De Poorter}, ``{End-to-End
  Learning From Spectrum Data: A Deep Learning Approach for Wireless Signal
  Identification in Spectrum Monitoring Applications},'' \emph{IEEE Access},
  vol.~6, pp. 18\,484--18\,501, 2018.

\bibitem{Rajendran2018}
S.~{Rajendran}, W.~{Meert}, D.~{Giustiniano}, V.~{Lenders}, and S.~{Pollin},
  ``{Deep Learning Models for Wireless Signal Classification With Distributed
  Low-Cost Spectrum Sensors},'' \emph{IEEE Transactions on Cognitive
  Communications and Networking}, vol.~4, no.~3, pp. 433--445, 2018.

\bibitem{Bu2020}
K.~{Bu}, Y.~{He}, X.~{Jing}, and J.~{Han}, ``{Adversarial Transfer Learning for
  Deep Learning Based Automatic Modulation Classification},'' \emph{IEEE Signal
  Processing Letters}, vol.~27, pp. 880--884, 2020.

\bibitem{Djolonga_etal_arXiv2020}
J.~Djolonga, J.~Yung, M.~Tschannen, and et~al., ``{On Robustness and
  Transferability of Convolutional Neural Networks},'' accessed: Feb. 18, 2021.
  [Online]. Available: https://arxiv.org/pdf/2007.08558.pdf.

\bibitem{DeepSigDataSet}
{DeepSig, Inc.} {RF Data Sets for Machine Learning: DeepSig Dataset RADIOML
  2018.01A}. Accessed Dec. 14, 2020. [Online]. Available:
  https://www.deepsig.ai/datasets.

\bibitem{MLChallenge}
{The CSP Blog}, ``{Data Set for the Machine Learning Challenge},'' accessed
  Dec. 7, 2020. [Online]. Available:
  https://cyclostationary.blog/2019/02/15/data-set-for-the-machine-learning-challenge.

\bibitem{Goodfellow-et-al-2016}
I.~Goodfellow, Y.~Bengio, and A.~Courville, \emph{Deep Learning}.\hskip 1em
  plus 0.5em minus 0.4em\relax MIT Press, 2016,
  https://www.deeplearningbook.org.

\end{thebibliography}
%
%
%

\end{document}